\documentclass[conference]{IEEEtran}
\IEEEoverridecommandlockouts
\usepackage{cite}
\usepackage{amsmath, amsthm, amssymb,amsfonts}
\usepackage{algorithmic}
\usepackage{graphicx}
\usepackage{textcomp}
\usepackage{tikz}
\usepackage{xcolor}
\def\BibTeX{{\rm B\kern-.05em{\sc i\kern-.025em b}\kern-.08em
    T\kern-.1667em\lower.7ex\hbox{E}\kern-.125emX}}
\usepackage{caption}
\usepackage{subcaption}
\usepackage[linesnumbered,ruled,vlined]{algorithm2e}
\SetKwInput{KwInput}{Input}                % Set the Input
\SetKwInput{KwOutput}{Output}              % set the Output

\usetikzlibrary{arrows,intersections}
\usetikzlibrary{mindmap,backgrounds}
\tikzstyle{block} = [draw, rectangle,
minimum height=3em, minimum width=5em, align=left]
\tikzstyle{pinstyle} = [pin edge={to-,thin,black}]    
    
\newtheorem{thm}{Theorem}[section]

\newtheorem{cor}[thm]{Corollary}
\newtheorem{lem}[thm]{Lemma}

\theoremstyle{definition}

\begin{document}

\title{Analytic Mutual Information in Bayesian Neural Networks%*\\
}

\author{\IEEEauthorblockN{Jae Oh Woo}
\IEEEauthorblockA{%\textit{dept. name of organization (of Aff.)} \\
\textit{Samsung SDS Research America}\\
San Jose, CA, USA \\
jaeoh.w@samsung.com}
}

\maketitle

\begin{abstract}
Bayesian neural networks have successfully designed and optimized a robust neural network model in many application problems, including uncertainty quantification. However, with its recent success, information-theoretic understanding about the Bayesian neural network is still at an early stage. Mutual information is an example of an uncertainty measure in a Bayesian neural network to quantify epistemic uncertainty. Still, no analytic formula is known to describe it, one of the fundamental information measures to understand the Bayesian deep learning framework. In this paper, we derive the analytical formula of the mutual information between model parameters and the predictive output by leveraging the notion of the point process entropy. Then, as an application, we discuss the parameter estimation of the Dirichlet distribution and show its practical application in the active learning uncertainty measures by demonstrating that our analytical formula can improve the performance of active learning further in practice.
\end{abstract}


\section{Introduction}
Uncertainty quantification plays a crucial role in managing and controlling exposed risks during optimization and decision-making in modern machine learning problems as the trained system is getting more complicated \cite{kendall2017uncertainties, jiang2018trust, begoli2019need, hullermeier2021aleatoric}. Bayesian approximation \cite{neal2012bayesian, srivastava2014dropout} and ensemble learning methods \cite{barber1998ensemble, chipman2007bayesian, pearce2020uncertainty} are two of the most widespread techniques to quantify uncertainties in the deep learning literature. The Bayesian neural network typically assumes a stochastic design to produce a posterior probability given prior knowledge. For example, the variational encoding approach is widely adopted \cite{Kingma2014}. The most straightforward Bayesian approximation is leveraging dropout layers \cite{gal2016dropout}. Another way is applying Laplace approximation \cite{ritter2018scalable, hobbhahn2020fast}. 

However, with its recent success, information-theoretic understanding about the Bayesian neural network is still at an early stage. Mutual information is an example of an uncertainty measure to quantify epistemic uncertainty \cite{matthies2007quantifying}. Another conditional entropy term is an example of an aleatoric uncertainty. Both uncertainty measures are practically crucial to evaluate the confidence or fairness of the model. For example, epistemic uncertainty captures the model uncertainty (lack of knowledge), and aleatoric uncertainty captures the inherent data uncertainty. Still, no analytic formulas are known to describe them \cite{houlsby2011bayesian, gal2016dropout}, which are fundamental information measures to understand the Bayesian deep learning framework.

In this paper, we derive the analytical formula of the mutual information between model parameters and the predictive output by leveraging the notion of the point process entropy \cite{baccelli2016entropy} and assuming that the intermediate encoded message in the Bayesian neural network follows a Dirichlet distribution since Dirichlet distribution family is the most natural and flexible family of probability distributions over a simplex in classification problem. Then, as a direct application, we discuss the parameter estimation of Dirichlet distribution and show its practical application in the active learning uncertainty measures by demonstrating that our analytical formula can improve the performance of active learning further in practice.  

\section{Information-Theoretic Formulation of Bayesian Neural Networks}\label{sec:sec2}
For simplicity, throughout this paper, we consider a classification problem with a Bayesian neural network approximated by Monte-Carlo (MC) dropouts \cite{srivastava2014dropout, gal2016dropout}. However, we note that our analytic framework does not have to be confined to the dropout regime. For example, our proposed framework can also be generalized to Gaussian process \cite{williams1998bayesian, rasmussen2006gaussianprocessesformachinelearning, milios2018dirichlet} or to leverage Laplace approximation in neural network \cite{kristiadi2020being, daxberger2021laplace}.

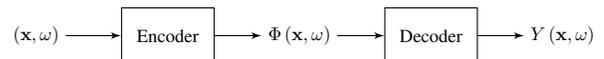
\begin{figure}[!htb]
\centering
\begin{tikzpicture}[scale=0.7, transform shape, auto,>=latex']
% We start by placing the blocks
    \node [name=message-source] {$\left(\mathbf{x},\omega \right)$};
    \node [block, name=encoder, right of=message-source, node distance=2.5cm] {Encoder};
    \node [name=channel, right of=encoder, node distance=2.5cm] { $\Phi\left(\mathbf{x},\omega \right)$};
    \node [block, name=decoder, right of=channel, node distance=2.5cm] {Decoder};
    \node [name=user, right of=decoder, node distance=2.5cm] {$Y\left(\mathbf{x},\omega \right)$};

    % Once the nodes are placed, connecting them is easy.
    \draw [->] (message-source) -- node[name=message, align=center] {} (encoder);
    \draw [->] (encoder) -- node[name=codeword, align=center] {} (channel);
    \draw [->] (channel) -- node[name=receivedword, align=center, text width=1.75cm,
                                 font=\scriptsize] {} (decoder); 
    \draw [->] (decoder) -- node[name=decodedmsg, align=center, text width=1.75cm] {} (user);
\end{tikzpicture}
\caption{\label{fig:bayesian_channel}Bayesian neural network channel framework}
\end{figure}

In an information-theoretic point of view, we can simplify the Bayesian neural network $\Phi\left(\cdot,\omega\right)$ with stochastic model parameters $\omega$ as an encoder-decoder communication channel. Given the data $\mathbf{x}$, the sender sends a message $\left(\mathbf{x},\omega \right)$ equipped with model parameters $\omega$ through the Bayesian channel, then the receiver receives a message  $Y\left(\mathbf{x},\omega\right)$ through the decoder. Figure \ref{fig:bayesian_channel} illustrates a diagram in this communication process.

% \begin{align*}
%     \left( \mathbf{x}, \omega \right) \to \Phi  \left( \mathbf{x}, \omega \right) := \left(P_1 \left( \mathbf{x}, \omega \right) ,\cdots, P_C \left( \mathbf{x}, \omega \right) \right) \to Y \left( \mathbf{x}, \omega \right) .
% \end{align*}
% The block diagram code is probably more verbose than necessary

Under this framework, the Bayesian deep neural network $\Phi$ produces the intermediate prediction probability for a data point $\mathbf{x}$:% is defined by
\begin{align*}
   {\Phi}\left(\mathbf{x}, \omega \right) :=\left(P_1(\mathbf{x},\omega), \cdots, P_C(\mathbf{x},\omega)\right) \in \Delta^C,
\end{align*}
where $\Delta^C = \{ (p_1,\cdots,p_C) : p_1+\cdots+p_C = 1, p_i\geq 0 \text{ for each }  i\}$ and $C$ is the number of classes. For the final class output $Y$, it is assumed to be a multinoulli distribution (or categorical distribution):
\begin{align*}
Y(\mathbf{x},\omega) := \begin{cases}
      1 & \text{with probability $P_1(\mathbf{x},\omega)$}\\
      \vdots & \vdots\\
      C & \text{with probability $P_C(\mathbf{x},\omega)$}.
    \end{cases}  
\end{align*}
Similar to find the channel capacity of the AWGN communication channel under power constraints \cite{shannon1948mathematical}, one may ask a similar question about the capacity of this Bayesian channel which is the mutual information between the model parameters $\omega$ and the output $Y$ denoting by $\mathfrak{I}\left( \omega, Y \left(\mathbf{x},\omega \right) \right)$ given $\mathbf{x}$, a.k.a. $\text{BALD}[\mathbf{x}]$ \cite{lindley1956measure, houlsby2011bayesian, gal2017deep}. In practice, controlling $\omega$ is not straightforward, but we can control the family of the encoded messages $\Phi\left(\mathbf{x},\omega \right)$ in a tractable manner \cite{gal2017deep, Kingma2014, tzikas2008variational}. Since $Y(\mathbf{x},\omega)$ only depends on $\Phi(\mathbf{x},\omega)$, by focusing on $\Phi\left(\mathbf{x},\omega \right)$, we may estimate the mutual information between the model parameters and the channel output \cite{gal2017deep}:
\begin{align}
    &\text{BALD}[\mathbf{x}] := \mathfrak{I}\left( \omega, Y\left(\mathbf{x},\omega\right) \right) \label{eq:bald1}\\
    =& H(Y\left(\mathbf{x},\omega\right))-\mathbb{E}_{\omega}\left[ H\left(Y\left(\mathbf{x},\omega\right) | \omega \right)\right]  \label{eq:bald2}\\
    =& H(Y\left(\mathbf{x},\omega\right)) - \mathbb{E}_{\Phi}\left[ H\left(Y\left(\mathbf{x},\omega\right) | \Phi\left(\mathbf{x},\omega\right) \right) \right] \label{eq:bald3}\\
    =&\mathfrak{I}\left( \Phi\left(\mathbf{x},\omega\right), Y(\mathbf{x},\omega) \right), \label{eq:bald4}
\end{align}
where $ H(Y\left(\mathbf{x},\omega\right))$ represents the Shannon entropy by marginalizing out the randomness of $\omega$ in $Y\left(\mathbf{x},\omega\right)$ and $\mathfrak{I}(\cdot,\cdot)$ represents a mutual information between two quantities. We remark that the equation \eqref{eq:bald3} is used to numerically estimate $\text{BALD}[\mathbf{x}]$ \cite{gal2017deep, KAJvAYGBatchBALD}.

The formulations of the mutual information \eqref{eq:bald1} - \eqref{eq:bald4} look natural, but we note that $\omega$ or $\Phi\left(\mathbf{x},\omega\right)$ is on a continuous domain, and $Y(\mathbf{x},\omega)$ is on a discrete domain. This combined domain implies that we cannot directly apply Shannon entropy and differential entropy notions. One immediate question is what the joint entropy between $\Phi\left(\mathbf{x},\omega\right)$ and $Y(\mathbf{x},\omega)$ is. Therefore, we first need to have a generalized notion of the entropy measures fitting into this Bayesian neural network framework.% to further understand the mutual information quantity.

By leveraging the point process entropy \cite{mcfadden1965entropy, fritz1973approach, papangelou1978entropy, daley2007introduction, baccelli2016entropy}, we can generalize the notion of the entropy in this combined domain. We note that a notion of the entropy for a discrete-continuous mixture can be applied in this Bayesian neural network \cite{nair2006entropy}. But the discrete-continuous mixture is a limited case of a point process, i.e., the point process entropy is a generalized definition of the discrete-continuous mixture entropy. Therefore, we keep the notion of the point process entropy in this paper. So equipping with the point process entropy, we need to consider a generalized notion of probability distribution on the combined domain, a.k.a. Jannosy density function \cite{daley2007introduction}. Following the usual point process entropy calculation, we may write a Janossy density function of $\left( \Phi\left(\mathbf{x},\omega\right), Y\left(\mathbf{x},\omega\right) \right)$ on $\Delta^C\times [C]$ as follows:
\begin{align}\label{eq:janossy}
    j\left( \mathbf{p}, y=i \right)= p_i f\left(\mathbf{p} \right),
\end{align}
where $\mathbf{p} := \left( p_1,\cdots, p_C \right)$ and $f(\cdot)$ is a density function of $\Phi\left(\mathbf{x},\omega\right)$. Then the joint entropy of $ \Phi\left(\mathbf{x},\omega\right)$ and $Y(\mathbf{x},\omega)$ can be defined as
\begin{align}\label{eq:joint_ent}
    &\mathfrak{H}\left( \Phi\left(\mathbf{x},\omega\right), Y\left(\mathbf{x},\omega\right) \right) \notag \\
    =& - \sum_{i=1}^{C}\int_{\Delta^c}  j\left( \mathbf{p}, y=i \right) \log  j\left( \mathbf{p}, y=i \right) \text{d}\mathbf{p}.
\end{align}
By plugging \eqref{eq:janossy} into \eqref{eq:joint_ent}, we can further drive the following identities:
\begin{align*}%\label{eq:identity}
    &\mathfrak{H}\left( \Phi\left(\mathbf{x},\omega\right), Y\left(\mathbf{x},\omega\right) \right) \notag \\
    =& H\left( Y\left(\mathbf{x},\omega\right) \right) + \mathbb{E}_Y\left[h\left( \Phi\left(\mathbf{x},\omega\right)  | Y\left(\mathbf{x},\omega\right) \right)\right] \\
    =& h\left( \Phi\left(\mathbf{x},\omega\right) \right) + \mathbb{E}_\Phi\left[H\left( Y\left(\mathbf{x},\omega\right)  | \Phi\left(\mathbf{x},\omega\right) \right)\right],
\end{align*}
where $h(\cdot)$ represents the usual differential entropy. Therefore we may further write equivalent forms of the mutual information as follows:
\begin{align}
    (4)&= h\left(\Phi\left(\mathbf{x},\omega\right)\right) + H\left(Y\left(\mathbf{x},\omega) \right) \right) - \mathfrak{H}\left( \Phi\left(\mathbf{x},\omega\right), Y(\mathbf{x},\omega) \right), \label{eq:bald5}\\
    &= h\left(\Phi\left(\mathbf{x},\omega\right)\right) - \mathbb{E}_Y\left[h\left( \Phi\left(\mathbf{x},\omega\right)  | Y\left(\mathbf{x},\omega\right) \right)\right]\notag%\\
%    &= H(Y\left(\mathbf{x},\omega\right)) - \mathbb{E}_{\Phi}\left[ H\left(Y(\omega) | \Phi(\omega) \right) \right].
\end{align}
%where $h(\cdot)$ is a usual differential entropy, and $\mathfrak{H}\left(\cdot,\cdot\right)$ is the generalized point process entropy.

Then, to establish the analytical formula, we assume that the distribution of $\Phi\left(\mathbf{x},\omega\right)$ follows Dirichlet distribution. In Bayesian model, the Gaussian-softmax-Dirichlet regime is a natural sequential application to generate a classification probability, by applying multivariate Gaussian to soft-max operation, then approximating it to Dirichlet distribution. Therefore, our choice of Dirichlet distribution is widely adopted in the literature of Monte-Carlo dropouts, Laplace approximation-based neural networks, and any Gaussian processes \cite{mackay1998choice, kristiadi2020being, williams1998bayesian, milios2018dirichlet, woo2021baba}. %We remark that Beta distribution is a marginal distribution of the Dirichlet distribution in each dimension. %In this dropout regime, the fitness of Beta distribution in the marginal of $\Phi\left(\mathbf{x},\omega\right)$ empirically supports the Dirichlet distribution assumption as shown in Figure \ref{fig:beta_approx} \cite{woo2021baba}:

\section{Main Results}
\begin{figure*}[t]
  \centering
  \begin{subfigure}{0.45\linewidth}
    \includegraphics[scale=0.275]{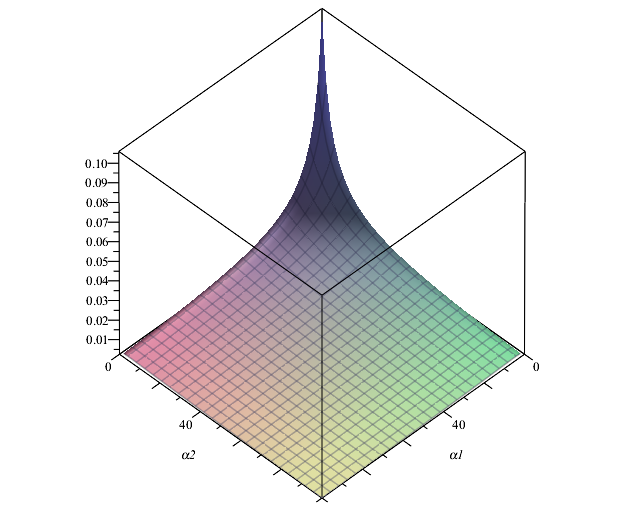}
       \caption{Analytic epistemic uncertainty (mutual information)}
   \label{fig:epistemic2d}
  \end{subfigure}
  \begin{subfigure}{0.45\linewidth}
    \includegraphics[scale=0.275]{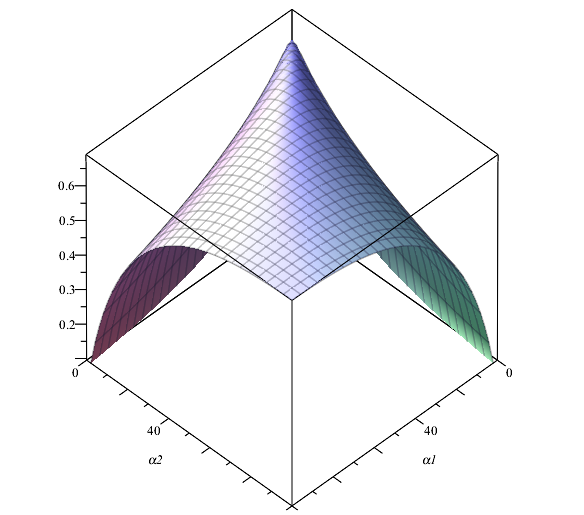}
       \caption{Analytic aleatoric uncertainty (conditional entropy)}
   \label{fig:aleatoric2d}
  \end{subfigure}
  \caption{Analytic uncertainties when $C=2$, i.e., $\Phi\left(\mathbf{x},\omega\right)\sim\text{Beta}\left(\alpha_1,\alpha_2\right)$ (=Beta distribution). Epistemic uncertainty tends to increase as $\alpha_1,\alpha_2\to 0$, and aleatoric uncertainty tends to increase as $\alpha_1,\alpha_2\to +\infty$.  }
\end{figure*}
In this section, we state our main results regarding the analytical form of the mutual information and its variant between model parameters $\omega$ and the predictive output $Y$ of Bayesian neural networks. The key assumption in our result is that the encoded message $\Phi\left(\mathbf{x},\omega\right)$ follows Dirichlet distribution with positive parameters $\left(\alpha_1,\cdots, \alpha_C\right)$. For the sake of brevity, let $\boldsymbol{\alpha}=\left(\alpha_1,\cdots, \alpha_C\right)$ and $\boldsymbol{\alpha}(i,++)=\left(\alpha_1,\cdots, \alpha_{i-1}, \alpha_i+1, \alpha_{i+1},\cdots, \alpha_C\right)$.

First we note that the entropy term can be decomposed into two uncertainty as below. The mutual information captures the epistemic uncertainty, and the conditional entropy captures the aleatoric uncertainty.
\begin{align}
    H(Y\left(\mathbf{x},\omega\right)) = \underbrace{\mathfrak{I} \left(\omega, Y \left(\mathbf{x},\omega\right)\right)}_\text{epistemic uncertainty}+ \underbrace{\mathbb{E}_\omega\left[H\left( Y \left(\mathbf{x},\omega\right) | \omega \right)\right]}_\text{aleatoric uncertainty}. \label{eq:uncertainty}
\end{align}
The epistemic uncertainty captures the model uncertainty (lack of knowledge), and the aleatoric uncertainty captures the data uncertainty \cite{matthies2007quantifying}. The decomposition \eqref{eq:uncertainty} implies the analytic formula of the aleatoric uncertainty as well. Our main results are Theorem \ref{thm:main1} and Corollary \ref{cor:aleatoric} for both uncertainties.

\begin{thm}\label{thm:main1}
Assume that $\Phi\left(\mathbf{x},\omega\right):=\left(P_1,\cdots, P_C\right)\sim \text{Dirichlet}(\alpha_1,\cdots, \alpha_C)$. Then the mutual information $\mathfrak{I}_{\text{Dirichlet}} \left( \omega, Y(\mathbf{x},\omega) \right)$ can be analytically calculated as follows.

{\small
\begin{align*}
    &\mathfrak{I}_{\text{Dirichlet}} \left( \omega, Y(\mathbf{x},\omega) \right) = \mathfrak{I}_{\text{Dirichlet}} \left( \Phi\left(\mathbf{x},\omega\right), Y(\mathbf{x},\omega) \right) \notag \\
    %=& h\left( \Phi\left(\mathbf{x},\omega\right) \right) + H(Y) - \mathfrak{H}_{\text{Dirichlet}} \left( \Phi\left(\mathbf{x},\omega\right), Y(\mathbf{x},\omega) \right)\\
    =& \left(\sum_{k=1}^{C} \alpha_k-C\right)\Psi\left(\sum_{k=1}^{C} \alpha_k\right)-\sum_{i=1}^{C}\left(\alpha_i-1\right)\Psi\left(\alpha_i\right)\notag\\
    -&\sum_{i=1}^{C}\left(  \frac{\alpha_i}{\sum_{k=1}^{C}\alpha_k} \right)\log\left(  \frac{\alpha_i}{\sum_{k=1}^{C}\alpha_k} \right)\notag\\
    +&\sum_{i=1}^{C}\sum_{j\neq i}\frac{\left(\alpha_j-1 \right)B\left(\boldsymbol{\alpha}(i,++)\right)}{B\left(\boldsymbol{\alpha}\right)} \left[ \Psi\left( \alpha_j\right)-\Psi\left(\left(\sum_{k=1}^{C} \alpha_k\right) +1\right) \right]\notag\\
    +&\sum_{i=1}^{C} \frac{\alpha_i B\left(\boldsymbol{\alpha}(i,++)\right)}{B\left(\boldsymbol{\alpha}\right)} \left[ \Psi\left( \alpha_i+1 \right)-\Psi\left(\left(\sum_{k=1}^{C} \alpha_k\right) +1\right) \right],
\end{align*}
}
where $B\left(\boldsymbol{\alpha}\right) = \frac{\Gamma(\alpha_1)\cdots\Gamma(\alpha_C)}{\Gamma\left(\sum_{k=1}^{C}\alpha_k\right)}$, $\Gamma(\cdot)$ is a Gamma function, and $\Psi(\cdot)$ is a Digamma function.
\end{thm}

\begin{cor}\label{cor:aleatoric}
Given the Bayesian neural network with $\Phi\left(\cdot,\omega\right)$, the aleatoric uncertainty can be analytically calculated as follows.
{\small
\begin{align*}
    &\text{Aleatoric uncertainty}\left[\mathbf{x}\right] :=\mathbb{E}_\omega\left[H\left( Y \left(\mathbf{x},\omega\right) | \omega \right)\right] \notag \\
    %=& h\left( \Phi\left(\mathbf{x},\omega\right) \right) + H(Y) - \mathfrak{H}_{\text{Dirichlet}} \left( \Phi\left(\mathbf{x},\omega\right), Y(\mathbf{x},\omega) \right)\\
    =& -\left(\sum_{k=1}^{C} \alpha_k-C\right)\Psi\left(\sum_{k=1}^{C} \alpha_k\right)+\sum_{i=1}^{C}\left(\alpha_i-1\right)\Psi\left(\alpha_i\right)\notag\\
    %-&\sum_{i=1}^{C}\left(  \frac{\alpha_i}{\sum_{k=1}^{C}\alpha_k} \right)\log\left(  \frac{\alpha_i}{\sum_{k=1}^{C}\alpha_k} \right)\notag\\
    -&\sum_{i=1}^{C}\sum_{j\neq i}\frac{\left(\alpha_j-1 \right)B\left(\boldsymbol{\alpha}(i,++)\right)}{B\left(\boldsymbol{\alpha}\right)} \left[ \Psi\left( \alpha_j\right)-\Psi\left(\left(\sum_{k=1}^{C} \alpha_k\right) +1\right) \right]\notag\\
    -&\sum_{i=1}^{C} \frac{\alpha_i B\left(\boldsymbol{\alpha}(i,++)\right)}{B\left(\boldsymbol{\alpha}\right)} \left[ \Psi\left( \alpha_i+1 \right)-\Psi\left(\left(\sum_{k=1}^{C} \alpha_k\right) +1\right) \right].
\end{align*}
}
\end{cor}

Figures \ref{fig:epistemic2d} and \ref{fig:aleatoric2d} illustrate the behavior of two uncertainties along parameters of Dirichlet distribution when $C=2$.

\section{Proof of Theorem \ref{thm:main1}} \label{proof:main1}
First, we note that the density function $f(\cdot)$ of $\text{Dirichlet}(\alpha_1,\cdots, \alpha_C)$ is given by
\begin{align}
    f\left( p_1,\cdots, p_C \right) = \frac{1}{B\left(\boldsymbol{\alpha}\right)}\prod_{i=1}^{C}p_i^{\alpha_i-1}.
\end{align}
Then to derive the analytical form, we shall calculate each term in the equation \eqref{eq:bald5}.
\begin{align*}
    &\mathfrak{I}_{\text{Dirichlet}} \left( \Phi\left(\mathbf{x},\omega\right), Y(\mathbf{x},\omega) \right)\\
    =& h\left( \Phi\left(\mathbf{x},\omega\right) \right) + H(Y\left(\mathbf{x},\omega\right)) - \mathfrak{H}_{\text{Dirichlet}} \left( \Phi\left(\mathbf{x},\omega\right), Y(\mathbf{x},\omega) \right). 
\end{align*}

Given $\Phi\left(\mathbf{x},\omega\right):=\left(P_1,\cdots, P_C\right)\sim \text{Dirichlet}(\alpha_1,\cdots, \alpha_C)$, the first differential entropy of Dirichlet distribution is well-known \cite{ebrahimi2011information, lin2016dirichlet}.
\begin{align}
    h\left( \Phi\left(\mathbf{x},\omega\right) \right) =& -\int_{\Delta^C }f\left( \mathbf{p}\right) \log  f\left( \mathbf{p} \right) \text{d}\mathbf{p}\notag\\
    =&\log B\left(\boldsymbol{\alpha}\right) + \left(\sum_{k=1}^{C} \alpha_k-C\right)\Psi\left(\sum_{k=1}^{C} \alpha_k\right)\notag\\
    &-\sum_{i=1}^{C}\left(\alpha_i-1\right)\Psi\left(\alpha_i\right).\label{eq:first_term}
\end{align}
For the second entropy term, we first need to use a simple property of Dirichlet distribution.
\begin{align}
    \mathbb{E} P_i = \frac{\alpha_i}{\sum_{k=1}^{C}\alpha_k}.\label{eq:diri_mean}
\end{align}
Then the second term can be obtained by following the Shannon entropy with the equation \eqref{eq:diri_mean}.
\begin{align}
     H(Y\left(\mathbf{x},\omega\right)) =& -\sum_{i=1}^{C} \mathbb{E}P_i\log \mathbb{E}P_i \notag \\
     =&-\sum_{i=1}^{C}\left(  \frac{\alpha_i}{\sum_{k=1}^{C}\alpha_k} \right)\log\left(  \frac{\alpha_i}{\sum_{k=1}^{C}\alpha_k} \right).\label{eq:second_term}
\end{align}

%Assume that $P\sim \text{Beta}(\alpha,\beta)$. Then we have a density function of $P$ as \begin{align*}
%f(x)=\frac{x^{\alpha-1}(1-x)^{\beta-1}}{B(\alpha,\beta)}.
%\end{align*}

For the third joint entropy term, we need to prove the following lemma.
\begin{lem}\label{lem:dirichlet2}
Assume that $\Phi\left(\mathbf{x},\omega\right):=\left(P_1,\cdots, P_C\right)\sim \text{Dirichlet}(\alpha_1,\cdots, \alpha_C)$.
{\small
\begin{align*}
    &\mathbb{E}\left[ P_i \log P_j \right] \\
    =& \begin{cases}
      \frac{B\left(\boldsymbol{\alpha}(i,++)\right)}{B\left(\boldsymbol{\alpha}\right)} \left[ \Psi\left( \alpha_i+1 \right)-\Psi\left(\left(\sum_{k=1}^{C} \alpha_k\right) +1\right) \right] & \text{if $i= j$,}\\
       \frac{B\left(\boldsymbol{\alpha}(i,++)\right)}{B\left(\boldsymbol{\alpha}\right)} \left[ \Psi\left( \alpha_j\right)-\Psi\left(\left(\sum_{k=1}^{C} \alpha_k\right) +1\right) \right]  & \text{if $i\neq j$}.
    \end{cases}  
\end{align*}
}
\end{lem}
To prove the Lemma \ref{lem:dirichlet2}, first we consider the $i=j$ case.
\begin{align*}
  &\mathbb{E}\left[ P_i \log P_i \right]= \frac{1}{B\left(\boldsymbol{\alpha}\right)} \int_{\Delta^C} \left( p_i\log p_i \right) \prod_{k=1}^{C}p_k^{\alpha_k-1} \text{d}\mathbf{p}\\
 % =&\frac{1}{B\left(\boldsymbol{\alpha}\right)} \int_{\Delta^C}  p_i^{\alpha_i}\log p_i  \prod_{k\neq i}p_k^{\alpha_k-1} \text{d}\mathbf{p}\\
  =&\frac{1}{B\left(\boldsymbol{\alpha}\right)} \int_{\Delta^C}  \frac{\text{d}}{\text{d}\alpha_i} p_i^{\alpha_i} \prod_{k\neq i}p_k^{\alpha_k-1} \text{d}\mathbf{p}\\
  %=&\frac{1}{B\left(\boldsymbol{\alpha}\right)}  \frac{\text{d}}{\text{d}\alpha_i} \int_{\Delta^C} p_i^{\alpha_i} \prod_{k\neq i}p_k^{\alpha_k-1} \text{d}\mathbf{p}\\
  =&\frac{1}{B\left(\boldsymbol{\alpha}\right)}  \frac{\text{d}}{\text{d}\alpha_i} B\left(\boldsymbol{\alpha}(i,++)\right)\\
  =& \frac{B\left(\boldsymbol{\alpha}(i,++)\right)}{B\left(\boldsymbol{\alpha}\right)} \left[ \Psi\left( \alpha_i+1 \right)-\Psi\left(\left(\sum_{k=1}^{C} \alpha_k\right) +1\right) \right].
\end{align*}
Note that we may interchange the differentiation and the integral operator by applying Lebesgue's dominated convergence theorem \cite{folland1999real}. The last equality can be derived by the definition of the Digamma function \cite{beukers2001special}. Similarly, for the $i\neq j$ case,
\begin{align*}
  &\mathbb{E}\left[ P_i \log P_j \right]= \frac{1}{B\left(\boldsymbol{\alpha}\right)} \int_{\Delta^C} \left( p_i\log p_j \right) \prod_{k=1}^{C}p_k^{\alpha_k-1} \text{d}\mathbf{p}\\
%  =&\frac{1}{B\left(\boldsymbol{\alpha}\right)} \int_{\Delta^C}  p_i^{\alpha_i}p_j^{\alpha_j-1}\log p_j  \prod_{k\neq i,j}p_k^{\alpha_k-1} \text{d}\mathbf{p}\\
  =&\frac{1}{B\left(\boldsymbol{\alpha}\right)} \int_{\Delta^C}   p_i^{\alpha_i} \frac{\text{d}}{\text{d}\alpha_j}p_j^{\alpha_j-1} \prod_{k\neq i,j}p_k^{\alpha_k-1} \text{d}\mathbf{p}\\
 % =&\frac{1}{B\left(\boldsymbol{\alpha}\right)}\frac{\text{d}}{\text{d}\alpha_j} \int_{\Delta^C}   p_i^{\alpha_i} p_j^{\alpha_j-1} \prod_{k\neq i}p_k^{\alpha_k-1} \text{d}\mathbf{p}\\
  =&\frac{1}{B\left(\boldsymbol{\alpha}\right)}  \frac{\text{d}}{\text{d}\alpha_j} B\left(\boldsymbol{\alpha}(i,++)\right)\\
  =& \frac{B\left(\boldsymbol{\alpha}(i,++)\right)}{B\left(\boldsymbol{\alpha}\right)} \left[ \Psi\left( \alpha_j\right)-\Psi\left(\left(\sum_{k=1}^{C} \alpha_k\right) +1\right) \right].
\end{align*}
Finally, we have the following identity by plugging the Janossy density of $\left( \Phi\left(\mathbf{x},\omega\right), Y(\mathbf{x},\omega)\right)$ into the equation \eqref{eq:joint_ent}:
\begin{align*}
    &\mathfrak{H}_{\text{Dirichlet}} \left( \Phi\left(\mathbf{x},\omega\right), Y(\mathbf{x},\omega) \right) \\
% =& - \sum_{i=1}^{C}\int_{\Delta^c}  j\left( \mathbf{p}, y=i \right) \log  j\left( \mathbf{p}, y=i \right) \text{d}\mathbf{p}\\
    =& \left(\log B\left(\boldsymbol{\alpha}\right)\right) \sum_{i=1}^{C}\mathbb{E} \left[ P_i \right] -\sum_{i=1}^{C}\sum_{j\neq i}\left(\alpha_j-1 \right)\mathbb{E}\left[ P_i \log P_j \right]\\
    &-\sum_{i=1}^{C}\alpha_i \mathbb{E}\left[ P_i \log P_i \right] =: (*).
\end{align*}
By applying Lemma \ref{lem:dirichlet2}, we have
{\small
\begin{align}
    (*)&= \log B\left(\boldsymbol{\alpha}\right)\notag\\
    -&\sum_{i=1}^{C}\sum_{j\neq i}\frac{\left(\alpha_j-1 \right)B\left(\boldsymbol{\alpha}(i,++)\right)}{B\left(\boldsymbol{\alpha}\right)} \left[ \Psi\left( \alpha_j\right)-\Psi\left(\left(\sum_{k=1}^{C} \alpha_k\right) +1\right) \right]\notag\\
    -&\sum_{i=1}^{C} \frac{\alpha_i B\left(\boldsymbol{\alpha}(i,++)\right)}{B\left(\boldsymbol{\alpha}\right)} \left[ \Psi\left( \alpha_i+1 \right)-\Psi\left(\left(\sum_{k=1}^{C} \alpha_k\right) +1\right) \right]. \label{eq:third_term}
\end{align}
}
By combining three terms \eqref{eq:first_term}, \eqref{eq:second_term}, and \eqref{eq:third_term} in the equation \eqref{eq:bald5}, Theorem \ref{thm:main1} follows.

\section{Parameter Estimation in Dirichlet Distribution}\label{sec:minka}
In Bayesian neural network with MC dropouts, $\Phi\left(\mathbf{x},\omega\right)$ is typically given as a collection of Monte-Carlo samples which are obtained from the Gaussian-softmax-Dirichlet regime as explained in Section \ref{sec:sec2}. Then given these samples, it is necessary to estimate appropriate parameters of the Dirichlet distribution to calculate the mutual information. In this section, we summarize the maximum-likelihood parameter estimation following Minka's approximation method \cite{Minka00estimatinga}.
\subsection{Minka's Fixed Point Iteration}
Given Monte-Carlo samples of $\Phi\left(\mathbf{x},\omega\right)$, let $ \mathbb{E}{p_k}$ be the sample mean of the $k$-th class probability. We have the following recurrent relation for the fixed-point iteration by taking the gradient in the likelihood to be zero.
\begin{align}\label{eq:fixedpoint}
    \Psi\left(\alpha_k\right) = \Psi\left(\sum_{k} \alpha_{k}\right) + \log \mathbb{E}{p_k}.
\end{align}
This implies the following iterative formula to find the fixed point.
\begin{align*}
    \alpha_k^{new} \leftarrow \Psi^{-1}\left[\Psi\left(\sum_{k} \alpha_{k}^{old}\right) + \log \mathbb{E}{p_k} \right].
\end{align*}
However, this iterative formula requires inverting the Digamma function $\Psi(x)$. We may leverage the Minka's asymptotic approximation of the inverse of the Digamma function \cite{Minka00estimatinga}:
\begin{align*}
    \Psi^{-1}(y) \approx & \begin{cases}
      \exp(y)+1/2 \text{ if $y\geq -2.22$},\\
      -\frac{1}{y+\gamma} \text{ if $y< -2.22$},
    \end{cases}  
\end{align*}
where $\gamma=-\Psi(1)$. We continue the iteration until it reaches a fixed point, but practically for batch tensor iteration, we fix a sufficiently large number of fixed-point iterations.

We remark that the initial choice of $\alpha_k^{initial}$ affects significantly to the final fixed point since the equation \eqref{eq:fixedpoint} does not take into account the sample variance of each marginal. To accommodate the second order (variance) information for each marginal, we apply the following initial condition:
\begin{align*}
    \alpha_k^{initial} = \frac{\left(\mathbb{E}{p_k}\right)^2-\mathbb{E}{p_k}\mathbb{E}{p_k^2}}{\mathbb{E}{p_k^2}-\left(\mathbb{E}{p_k}\right)^2}.
\end{align*}
Finally, we note that when $\mathbb{E}{p_k^2}\ll \mathbb{E}{p_k}\approx 0$, we allow a degenerate Dirichlet distribution $\Phi\left(\omega,\mathbf{x}\right)$ by assuming $\alpha_k=0$ for numerical stability.

\section{Application in Active Learning}
\begin{figure*}[t]\label{fig:activelearning}
  \centering
  \begin{subfigure}{0.45\linewidth}
    \includegraphics[scale=0.25]{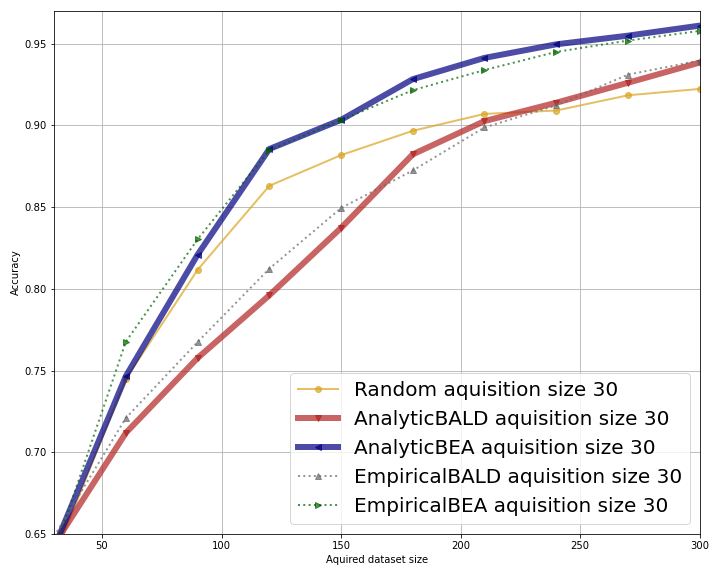}
     \caption{MNIST}
   \label{fig:mnist}
  \end{subfigure}
  \begin{subfigure}{0.45\linewidth}
    \includegraphics[scale=0.25]{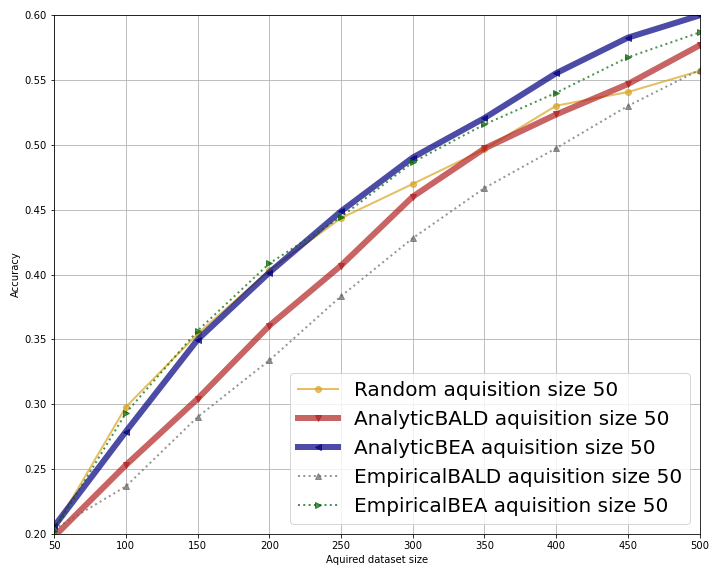}
    \caption{EMNIST}
   \label{fig:emnist}
  \end{subfigure}
  \caption{Active learning curves obtained from MNIST and EMNIST. Each curve is averaged after $3$ repeated experiments with different random seeds. As iteration proceeds, we observe that analytic formula can improve the performance of the active learning.}
  \label{fig:activelearning}
\end{figure*}
This section demonstrates the application of the derived analytic formula of the mutual information by comparing it with the numerically calculated quantity through active learning. In many application problems, labeling data by humans becomes very expensive as the dataset size grows. So in practice, it is critical to efficiently build a model by minimizing the efforts of human labeling from the unlabeled training data pool. To achieve this goal, we can apply the active learning approach \cite{balcan2010true, bangert2022medical, bangertactive, hao2021highly}. In active learning, we iteratively increment the training data from the unlabeled training pool and re-train the model. At each iteration, we typically use an uncertainty measure to select the most informative data points given the pre-defined incremental size, denoting it by $K$ in selecting the following training dataset up to the total active learning budget $K^{tot}$ which is the total number of labeled data points. Algorithm \ref{algo:activelearning} describes the general procedures of active learning. For the details of active learning, we recommend referring to articles \cite{cohn1996active, settles2009active, aggarwal2014active, gal2017deep, KAJvAYGBatchBALD, woo2021baba}. We list up mutual-information-related uncertainty measures for active learning under the Bayesian deep learning framework for our demonstration.
\begin{itemize}
    \item[1.] \textbf{Random}: $\text{Rand}[\mathbf{x}]:=U(\omega')$
    where $U(\cdot)$ is a uniform distribution which is independent to $\omega$. Random acquisition function assigns a random uniform value on $[0,1]$ to each data point. Random acquisition function is used for building a baseline accuracy.
    \item[2.] \textbf{BALD} (Bayesian active learning by disagreement) \cite{lindley1956measure, houlsby2011bayesian, gal2017deep}: $\text{BALD}[\mathbf{x}]:=\mathfrak{I}\left(\omega, Y\left( \mathbf{x}, \omega \right) \right)$. BALD is a mutual information to capture the epistemic uncertainty.
    \item[3.] \textbf{BalEntAcq} \cite{woo2021baba}:   
    $\text{BEA}[\mathbf{x}]:=
    \small
    \begin{cases}
        \frac{\text{BALD}[\mathbf{x}]}{ \text{MJEnt}[\mathbf{x}]} & \text{if $\text{MJEnt}[\mathbf{x}] \geq 0$},\\
        \frac{ \text{MJEnt}[\mathbf{x}]}{\text{BALD}[\mathbf{x}]} & \text{if $\text{MJEnt}[\mathbf{x}] < 0$},
    \end{cases}$
    where $\text{MJEnt}[\mathbf{x}]=\sum_i  \left( \mathbb{E}P_i \right) \left[ h(P_i^+) - \log \left( \mathbb{E}P_i \right) \right]$ and $P_i^+$ is the conjugate Beta posterior entropy of $P_i$ which follows $P_i^+ \sim \text{Beta}\left(\alpha'_i+1, \sum_{j\neq i} \alpha_j \right)$ in $\Phi\left( \mathbf{x},\omega \right)$. In many scenarios with Bayesian neural networks, BalEntAcq (BEA) has shown a superior performance \cite{woo2021baba}. %We also note that we apply the analytic formula only for $\text{BALD}[\mathbf{x}]$ in BABA for improving diversification. %\
\end{itemize} 

In our experiments, we use MNIST and EMINST datasets \cite{lecun-mnisthandwrittendigit-2010, cohen2017emnist}. MNIST is the most popular dataset to validate the performance of image-based deep learning models and the EMNIST dataset is a set of handwritten character digits aligning with the MNIST dataset. For the empirical BALD/or BEA, we apply the equation \eqref{eq:bald3}. For the analytic BALD/or BEA, we apply Theorem \ref{thm:main1}.

\begin{algorithm}
\small
\DontPrintSemicolon
  \textbf{Input:} 1) Total training dataset $\mathcal{D}_{\textbf{pool}}$, 2) randomly selected initial dataset $\mathcal{D}^{(0)}_{\textbf{training}}$, 3) $M$ as the number of dropout samples, 4) active learning budget $K$ for each iteration, 5) total active learning budget $K^{tot}$
  
  \textbf{Initialize} Bayesian neural network $\Phi$ and set $n\leftarrow 0$
  
  \textbf{Repeat} at iteration $n\geq 0$
  
  \quad Train the model $\Phi$ with $\mathcal{D}^{(n)}_{\textbf{training}}$
   
  \quad For each $\mathbf{x}\in \mathcal{D}_{\textbf{pool}} \setminus \mathcal{D}^{(n)}_{\textbf{training}}$, 
   
  \qquad Generate $M$ Dirichlet samples from $\Phi\left(\mathbf{x},\omega\right)$
   
    \qquad Estimate Drichlet parameters $\left(\alpha_1, \cdots, \alpha_C\right)$ following Minka's fixed point iteration %(and estimate Beta parameters $\left(\alpha'_i,\beta'_i\right)$ for each marginal in BABA)
    
    \qquad Calculate the uncertainty value
    
    \quad Set $\mathcal{D}^{(n+1)}_{\textbf{training}} \leftarrow$ $\mathcal{D}^{(n)}_{\textbf{training}} \bigcup $ $\left\{ \text{top } K \text{ uncertainty-valued } \mathbf{x}\in \mathcal{D}_{\textbf{pool}} \setminus \mathcal{D}^{(n)}_{\textbf{training}} \right\}$, and $n\leftarrow n+1$

  \textbf{Until} $\left| \mathcal{D}^{(n-1)}_{\textbf{training}}\right|$ reaches to $K^{tot}$
\caption{Active learning algorithm}\label{algo:activelearning}
\end{algorithm}

Figure \ref{fig:activelearning} shows the performance of active learning results with uncertainty measures. For the MNIST dataset, we use the acquisition size $K=30$ in each iteration up to $K^{tot}=300$ number of images. For EMNIST dataset, we use the acquisition size of $K=50$ for each iteration up to $K^{tot}=500$ number of images. We start from a randomly selected initial training set for both cases to train the initial model. We note that BALD suffers from improving the accuracy compared to the random case since it cannot effectively remove the redundancy. We confirm that analytic BALD and analytic BEA show similar in MNIST or better behavior in EMNIST with empirical BALD and empirical BEA. % similarly observed in BatchBALD \cite{KAJvAYGBatchBALD}.%We visually confirm that analytic BALD and analytic BABA show similar in MNIST or better behavior in EMNIST with empirical BALD and empirical BABA, implying that our derived analytical formula is aligned with the numerical calculation.  %We observe the same phenomenon in both empirical and analytical BALD cases.
% \subsection{MNIST}
% \begin{figure}[!htb]
%     \centering
%     \includegraphics[scale=0.3]{mnist.png}
%       \caption{Active learning curve for MNIST}
%   \label{fig:mnist}
% \end{figure}
% \subsection{EMNIST}
% \begin{figure}[!htb]
%     \centering
%     \includegraphics[scale=0.3]{emnist.png}
%       \caption{Active learning curve for EMNIST}
%   \label{fig:mnist}
% \end{figure}

\section{Conclusion}
This paper presented analytic mutual information in Bayesian neural networks and their application in active learning. We derived the analytical formula of the mutual information as an epistemic uncertainty or the conditional entropy as an aleatoric uncertainty. Aligning with the recent success of BalEngAcq (BEA) \cite{woo2021baba}, we expect that our analytical framework would enhance the understanding of BalEngAcq as well as the further applications of the Bayesian neural network to build a robust and reliable neural network model.

% \section*{Acknowledgment}
% The author thanks the helpful discussion with and suggestions from Patrick Bangert, Hankyu Moon, Sima Didari, and Heng Hao in Samsung SDS Research America.
\newpage

\bibliographystyle{IEEEtran} 
\bibliography{bibbib}

\begin{thebibliography}{10}
\providecommand{\url}[1]{#1}
\csname url@samestyle\endcsname
\providecommand{\newblock}{\relax}
\providecommand{\bibinfo}[2]{#2}
\providecommand{\BIBentrySTDinterwordspacing}{\spaceskip=0pt\relax}
\providecommand{\BIBentryALTinterwordstretchfactor}{4}
\providecommand{\BIBentryALTinterwordspacing}{\spaceskip=\fontdimen2\font plus
\BIBentryALTinterwordstretchfactor\fontdimen3\font minus
  \fontdimen4\font\relax}
\providecommand{\BIBforeignlanguage}[2]{{%
\expandafter\ifx\csname l@#1\endcsname\relax
\typeout{** WARNING: IEEEtran.bst: No hyphenation pattern has been}%
\typeout{** loaded for the language `#1'. Using the pattern for}%
\typeout{** the default language instead.}%
\else
\language=\csname l@#1\endcsname
\fi
#2}}
\providecommand{\BIBdecl}{\relax}
\BIBdecl

\bibitem{kendall2017uncertainties}
A.~Kendall and Y.~Gal, ``What uncertainties do we need in bayesian deep
  learning for computer vision?'' \emph{Advances in neural information
  processing systems}, vol.~30, 2017.

\bibitem{jiang2018trust}
H.~Jiang, B.~Kim, M.~Y. Guan, and M.~Gupta, ``To trust or not to trust a
  classifier,'' \emph{Neural Information Processing Systems}, 2018.

\bibitem{begoli2019need}
E.~Begoli, T.~Bhattacharya, and D.~Kusnezov, ``The need for uncertainty
  quantification in machine-assisted medical decision making,'' \emph{Nature
  Machine Intelligence}, vol.~1, no.~1, pp. 20--23, 2019.

\bibitem{hullermeier2021aleatoric}
E.~H{\"u}llermeier and W.~Waegeman, ``Aleatoric and epistemic uncertainty in
  machine learning: An introduction to concepts and methods,'' \emph{Machine
  Learning}, vol. 110, no.~3, pp. 457--506, 2021.

\bibitem{neal2012bayesian}
R.~M. Neal, \emph{Bayesian learning for neural networks}.\hskip 1em plus 0.5em
  minus 0.4em\relax Springer Science \& Business Media, 2012, vol. 118.

\bibitem{srivastava2014dropout}
N.~Srivastava, G.~Hinton, A.~Krizhevsky, I.~Sutskever, and R.~Salakhutdinov,
  ``Dropout: a simple way to prevent neural networks from overfitting,''
  \emph{The journal of machine learning research}, vol.~15, no.~1, pp.
  1929--1958, 2014.

\bibitem{barber1998ensemble}
D.~Barber and C.~M. Bishop, ``Ensemble learning in bayesian neural networks,''
  \emph{Nato ASI Series F Computer and Systems Sciences}, vol. 168, pp.
  215--238, 1998.

\bibitem{chipman2007bayesian}
H.~A. Chipman, E.~I. George, and R.~E. McCulloch, ``Bayesian ensemble
  learning,'' \emph{Advances in neural information processing systems},
  vol.~19, p. 265, 2007.

\bibitem{pearce2020uncertainty}
T.~Pearce, F.~Leibfried, and A.~Brintrup, ``Uncertainty in neural networks:
  Approximately bayesian ensembling,'' in \emph{International conference on
  artificial intelligence and statistics}.\hskip 1em plus 0.5em minus
  0.4em\relax PMLR, 2020, pp. 234--244.

\bibitem{Kingma2014}
D.~P. Kingma and M.~Welling, ``{Auto-Encoding Variational Bayes},'' in
  \emph{2nd International Conference on Learning Representations, {ICLR} 2014,
  Banff, AB, Canada, April 14-16, 2014, Conference Track Proceedings}, 2014.

\bibitem{gal2016dropout}
Y.~Gal and Z.~Ghahramani, ``Dropout as a bayesian approximation: Representing
  model uncertainty in deep learning,'' in \emph{international conference on
  machine learning}.\hskip 1em plus 0.5em minus 0.4em\relax PMLR, 2016, pp.
  1050--1059.

\bibitem{ritter2018scalable}
H.~Ritter, A.~Botev, and D.~Barber, ``A scalable laplace approximation for
  neural networks,'' in \emph{6th International Conference on Learning
  Representations, ICLR 2018-Conference Track Proceedings}, vol.~6.\hskip 1em
  plus 0.5em minus 0.4em\relax International Conference on Representation
  Learning, 2018.

\bibitem{hobbhahn2020fast}
M.~Hobbhahn, A.~Kristiadi, and P.~Hennig, ``Fast predictive uncertainty for
  classification with bayesian deep networks,'' \emph{arXiv preprint
  arXiv:2003.01227}, 2020.

\bibitem{matthies2007quantifying}
H.~G. Matthies, ``Quantifying uncertainty: modern computational representation
  of probability and applications,'' in \emph{Extreme man-made and natural
  hazards in dynamics of structures}.\hskip 1em plus 0.5em minus 0.4em\relax
  Springer, 2007, pp. 105--135.

\bibitem{houlsby2011bayesian}
N.~Houlsby, F.~Husz{\'a}r, Z.~Ghahramani, and M.~Lengyel, ``Bayesian active
  learning for classification and preference learning,'' \emph{arXiv preprint
  arXiv:1112.5745}, 2011.

\bibitem{baccelli2016entropy}
F.~Baccelli and J.~O. Woo, ``On the entropy and mutual information of point
  processes,'' in \emph{2016 IEEE International Symposium on Information Theory
  (ISIT)}.\hskip 1em plus 0.5em minus 0.4em\relax IEEE, 2016, pp. 695--699.

\bibitem{williams1998bayesian}
C.~K. Williams and D.~Barber, ``Bayesian classification with gaussian
  processes,'' \emph{IEEE Transactions on Pattern Analysis and Machine
  Intelligence}, vol.~20, no.~12, pp. 1342--1351, 1998.

\bibitem{rasmussen2006gaussianprocessesformachinelearning}
C.~Rasmussen and C.~Williams, ``Gaussian processes for machine learning.
  adaptive computation and machine learning,'' 2006.

\bibitem{milios2018dirichlet}
D.~Milios, R.~Camoriano, P.~Michiardi, L.~Rosasco, and M.~Filippone,
  ``Dirichlet-based gaussian processes for large-scale calibrated
  classification,'' \emph{arXiv preprint arXiv:1805.10915}, 2018.

\bibitem{kristiadi2020being}
A.~Kristiadi, M.~Hein, and P.~Hennig, ``Being bayesian, even just a bit, fixes
  overconfidence in relu networks,'' in \emph{International Conference on
  Machine Learning}.\hskip 1em plus 0.5em minus 0.4em\relax PMLR, 2020, pp.
  5436--5446.

\bibitem{daxberger2021laplace}
E.~Daxberger, A.~Kristiadi, A.~Immer, R.~Eschenhagen, M.~Bauer, and P.~Hennig,
  ``Laplace redux-effortless bayesian deep learning,'' \emph{Advances in Neural
  Information Processing Systems}, vol.~34, 2021.

\bibitem{shannon1948mathematical}
C.~E. Shannon, ``A mathematical theory of communication,'' \emph{The Bell
  system technical journal}, vol.~27, no.~3, pp. 379--423, 1948.

\bibitem{lindley1956measure}
D.~V. Lindley, ``On a measure of the information provided by an experiment,''
  \emph{The Annals of Mathematical Statistics}, pp. 986--1005, 1956.

\bibitem{gal2017deep}
Y.~Gal, R.~Islam, and Z.~Ghahramani, ``Deep bayesian active learning with image
  data,'' in \emph{International Conference on Machine Learning}.\hskip 1em
  plus 0.5em minus 0.4em\relax PMLR, 2017, pp. 1183--1192.

\bibitem{tzikas2008variational}
D.~G. Tzikas, A.~C. Likas, and N.~P. Galatsanos, ``The variational
  approximation for bayesian inference,'' \emph{IEEE Signal Processing
  Magazine}, vol.~25, no.~6, pp. 131--146, 2008.

\bibitem{KAJvAYGBatchBALD}
A.~Kirsch, J.~van Amersfoort, and Y.~Gal, ``Batchbald: Efficient and diverse
  batch acquisition for deep bayesian active learning,'' 2019.

\bibitem{mcfadden1965entropy}
J.~McFadden, ``The entropy of a point process,'' \emph{Journal of the Society
  for Industrial \& Applied Mathematics}, vol.~13, no.~4, pp. 988--994, 1965.

\bibitem{fritz1973approach}
J.~Fritz, ``An approach to the entropy of point processes,'' \emph{Periodica
  Mathematica Hungarica}, vol.~3, no. 1-2, pp. 73--83, 1973.

\bibitem{papangelou1978entropy}
F.~Papangelou, ``On the entropy rate of stationary point processes and its
  discrete approximation,'' \emph{Probability Theory and Related Fields},
  vol.~44, no.~3, pp. 191--211, 1978.

\bibitem{daley2007introduction}
D.~J. Daley and D.~Vere-Jones, \emph{An introduction to the theory of point
  processes: volume II: general theory and structure}.\hskip 1em plus 0.5em
  minus 0.4em\relax Springer Science \& Business Media, 2007, vol.~2.

\bibitem{nair2006entropy}
C.~Nair, B.~Prabhakar, and D.~Shah, ``On entropy for mixtures of discrete and
  continuous variables,'' \emph{arXiv preprint cs/0607075}, 2006.

\bibitem{mackay1998choice}
D.~J. MacKay, ``Choice of basis for laplace approximation,'' \emph{Machine
  learning}, vol.~33, no.~1, pp. 77--86, 1998.

\bibitem{woo2021baba}
J.~O. Woo, ``Active learning in bayesian neural networks: Balanced entropy
  learning principle,'' \emph{arXiv preprint arXiv:2105.14559}, 2021.

\bibitem{ebrahimi2011information}
N.~Ebrahimi, E.~S. Soofi, and S.~Zhao, ``Information measures of dirichlet
  distribution with applications,'' \emph{Applied Stochastic Models in Business
  and Industry}, vol.~27, no.~2, pp. 131--150, 2011.

\bibitem{lin2016dirichlet}
J.~Lin, ``On the dirichlet distribution,'' \emph{Master's Thesis}, 2016.

\bibitem{folland1999real}
G.~B. Folland, \emph{Real analysis: modern techniques and their
  applications}.\hskip 1em plus 0.5em minus 0.4em\relax John Wiley \& Sons,
  1999, vol.~40.

\bibitem{beukers2001special}
F.~Beukers, ``Special functions (encyclopedia of mathematics and its
  applications 71),'' \emph{Bulletin of the London Mathematical Society},
  vol.~33, no.~1, pp. 116--127, 2001.

\bibitem{Minka00estimatinga}
T.~P. Minka, ``Estimating a dirichlet distribution,'' Tech. Rep., 2000.

\bibitem{balcan2010true}
M.-F. Balcan, S.~Hanneke, and J.~W. Vaughan, ``The true sample complexity of
  active learning,'' \emph{Machine learning}, vol.~80, no.~2, pp. 111--139,
  2010.

\bibitem{bangert2022medical}
P.~Bangert, H.~Moon, J.~O. Woo, S.~Didari, and H.~Hao, ``Medical image labeling
  via active learning is 90\% effective,'' in \emph{Future of Information and
  Communication Conference}.\hskip 1em plus 0.5em minus 0.4em\relax Springer,
  2022, pp. 291--310.

\bibitem{bangertactive}
------, ``Active learning performance in labeling radiology images is 90\%
  effective,'' \emph{Frontiers in Radiology}, p.~13.

\bibitem{hao2021highly}
H.~Hao, H.~Moon, S.~Didari, J.~O. Woo, and P.~Bangert, ``Highly efficient
  representation and active learning framework for imbalanced data and its
  application to covid-19 x-ray classification,'' \emph{NeurIPS Data-Centric AI
  Workshop}, 2021.

\bibitem{cohn1996active}
D.~A. Cohn, Z.~Ghahramani, and M.~I. Jordan, ``Active learning with statistical
  models,'' \emph{Journal of artificial intelligence research}, vol.~4, pp.
  129--145, 1996.

\bibitem{settles2009active}
B.~Settles, ``Active learning literature survey,'' 2009.

\bibitem{aggarwal2014active}
C.~C. Aggarwal, X.~Kong, Q.~Gu, J.~Han, and S.~Y. Philip, ``Active learning: A
  survey,'' in \emph{Data Classification}.\hskip 1em plus 0.5em minus
  0.4em\relax Chapman and Hall/CRC, 2014, pp. 599--634.

\bibitem{lecun-mnisthandwrittendigit-2010}
\BIBentryALTinterwordspacing
Y.~LeCun and C.~Cortes, ``{MNIST} handwritten digit database,'' 2010. [Online].
  Available: \url{http://yann.lecun.com/exdb/mnist/}
\BIBentrySTDinterwordspacing

\bibitem{cohen2017emnist}
G.~Cohen, S.~Afshar, J.~Tapson, and A.~Van~Schaik, ``Emnist: Extending mnist to
  handwritten letters,'' in \emph{2017 International Joint Conference on Neural
  Networks (IJCNN)}.\hskip 1em plus 0.5em minus 0.4em\relax IEEE, 2017, pp.
  2921--2926.

\end{thebibliography}

% \bibitem{b1} G. Eason, B. Noble, and I. N. Sneddon, ``On certain integrals of Lipschitz-Hankel type involving products of Bessel functions,'' Phil. Trans. Roy. Soc. London, vol. A247, pp. 529--551, April 1955.
% \bibitem{b2} J. Clerk Maxwell, A Treatise on Electricity and Magnetism, 3rd ed., vol. 2. Oxford: Clarendon, 1892, pp.68--73.
% \bibitem{b3} I. S. Jacobs and C. P. Bean, ``Fine particles, thin films and exchange anisotropy,'' in Magnetism, vol. III, G. T. Rado and H. Suhl, Eds. New York: Academic, 1963, pp. 271--350.
% \bibitem{b4} K. Elissa, ``Title of paper if known,'' unpublished.
% \bibitem{b5} R. Nicole, ``Title of paper with only first word capitalized,'' J. Name Stand. Abbrev., in press.
% \bibitem{b6} Y. Yorozu, M. Hirano, K. Oka, and Y. Tagawa, ``Electron spectroscopy studies on magneto-optical media and plastic substrate interface,'' IEEE Transl. J. Magn. Japan, vol. 2, pp. 740--741, August 1987 [Digests 9th Annual Conf. Magnetics Japan, p. 301, 1982].
% \bibitem{b7} M. Young, The Technical Writer's Handbook. Mill Valley, CA: University Science, 1989.
% \end{thebibliography}

\end{document}